# Secure and Energy-Efficient Data Aggregation in Wireless Sensor Networks

Jaydip Sen

Member, ACM
Kolkata, INDIA
e-mail : jaydip.sen@acm.org

*Abstract*— **Data aggregation in intermediate nodes (called aggregator nodes) is an effective approach for optimizing consumption of scarce resources like bandwidth and energy in Wireless Sensor Networks (WSNs). However, in-network processing poses a problem for the privacy of the sensor data since individual data of sensor nodes need to be known to the aggregator node before the aggregation process can be carried out. In applications of WSNs, privacy-preserving data aggregation has become an important requirement due to sensitive nature of the sensor data. Researchers have proposed a number of protocols and schemes for this purpose. He et al. (INFOCOM 2007) have proposed a protocol - called CPDA - for carrying out additive data aggregation in a privacy-preserving manner for application in WSNs. The scheme has been quite popular and well-known. In spite of the popularity of this protocol, it has been found that the protocol is vulnerable to attack and it is also not energy-efficient. In this paper, we first present a brief state of the art survey on the current privacy-preserving data aggregation protocols for WSNS. Then we describe the CPDA protocol and identify its security vulnerability. Finally, we demonstrate how the protocol can be made secure and energy efficient.**

## I. INTRODUCTION *(HEADING 1)*

In recent years, wireless sensor networks (WSNs) have drawn considerable attention from the research community on issues ranging from theoretical research to practical applications. Special characteristics of WSNs, such as resource constraints on energy and computational power and security have been well-defined and widely studied [1, 2]. What has received less attention, however, is the critical privacy concern on information being collected, transmitted, and analyzed in a WSN. Such private and sensitive information may include payload data collected by sensors and transmitted through the network to a centralized data processing server. For example, a patient's blood pressure, sugar level and other vital signs are usually of critical privacy concern when monitored by a medical WSN which transmits the data to a remote hospital or doctor's office. Privacy concerns may also arise beyond data content and may focus on context information such as the location of a sensor initiating data communication. Effective countermeasure against the disclosure of both data and context-oriented private information is an indispensable prerequisite for deployment of WSNs in real-world applications.

Privacy protection has been extensively studied in various fields related to WSNs such as wired and wireless networking, databases and data mining. Nonetheless, the following inherent features of WSNs introduce unique challenges for privacy preservation in WSNs, and prevent the existing techniques from being directly transplanted: (i) *Uncontrollable environment*: Sensors may have to be deployed to an environment uncontrollable by the defender, such as a battlefield, enabling an adversary to launch physical attacks to capture sensor nodes or deploy counterfeit ones. As a result, an adversary may retrieve private keys used for secure communication and decrypt any communication eavesdropped by the adversary. (ii) *Sensor-node resource constraints*: battery-powered sensor nodes generally have severe constraints on their ability to store, process, and transmit the sensed data. As a result, the computational complexity and resource consumption of public-key ciphers is usually considered unsuitable for WSNs. (iii) *Topological constraints*: the limited communication range of sensor nodes in a WSN requires multiple hops in order to transmit data from the source to the base station. Such a multi-hop scheme demands different nodes to take diverse traffic loads. In particular, a node closer to the base station (i.e., data collecting and processing server) has to relay data from nodes further away from base station in addition to transmitting its own generated data, leading to higher transmission rate. Such an unbalanced network traffic pattern brings significant challenges to the protection of context-oriented privacy information. Particularly, if an adversary holds the ability of global traffic analysis, observing the traffic patterns of different nodes over the whole network, it can easily identify the sink and compromise context privacy, or even manipulate the sink node to impede the proper functioning of the WSN.

The unique challenges for privacy preservation in WSNs call for the development of effective privacy-preserving techniques. Supporting efficient in-network data aggregation while preserving data privacy has emerged as an important requirement in numerous wireless sensor network applications [3, 4, 5, 6, 7]. As a key approach to fulfilling this requirement of private data aggregation, *concealed data aggregation* (CDA) schemes have been proposed in which multiple source nodes send encrypted data to a sink along a converge-cast tree with aggregation of cipher-text being performed over the route [4, 5, 6, 7, 8, 9].

He et al. have proposed a *cluster-based private data aggregation* (CPDA) scheme in which the sensor nodes are randomly distributed into clusters [3]. The cluster leaders are responsible for directly aggregating data from the cluster members, with the communication secured by a shared key between a pair of communicating nodes. The aggregate function leverages algebraic properties of the polynomials to compute the desired aggregate value in a cluster. While the aggregation is carried out at the aggregator node in each cluster, it is guaranteed that no individual node gets to know the sensitive private values of other nodes in the cluster. The intermediate aggregate value in each cluster is further aggregated along the routing tree as the data packets move to the sink node. The privacy goal of the scheme is two-fold. First, the privacy of data has to be guaranteed end-to-end. While only the sink could learn about the final aggregation result, each node will have information of its own data and does not have any information about the data of other nodes. Second, to reduce the communication overhead, the data from different source nodes have to be efficiently combined by intermediate nodes (i.e. aggregation) along the path. Nevertheless, these intermediate nodes should not learn any information about the individual nodes' data. The authors of the CPDA scheme have presented performance results of the protocol to demonstrate the efficiency and security of the protocol. The CPDA protocol has become quite popular, and to the best of our knowledge, there has been no identified vulnerability of the protocol published in the literature.

In this paper, we first provide an alternative approach to CPDA protocol so that sensor data aggregation can be carried out in a more efficient manner and then proceed to describe a potential attack on CPDA to show the vulnerability of the scheme. We also propose necessary modifications in the scheme to defend against the identified vulnerability.

The rest of this paper is organized as follows. Section II briefly describes some of the existing secure and privacy-preserving aggregation schemes for WSNs. Section III provides a brief discussion on the CPDA scheme. Section IV presents an attack on the CPDA scheme. Section V discusses how the CPDA scheme can be modified to make it more secure and computationally more efficient. Finally, Section VI concludes the paper while highlighting some future scope of work.

## II. RELATED WORK

Researchers have considered security as an important requirement in designing any aggregation scheme for WSNs. In addition, privacy also has been identified as one of the emerging requirements in many WSNs applications, and hence attracted a lot of attention.

Hu and Evans have proposed a *secure aggregation* (SA) protocol that uses μTESLA protocol [10]. The protocol is resilient to both intruder devices and single device key compromises. The scheme uses a tree structure to represent the nodes, where the leaves represent the sensor nodes and the nodes in the higher levels act at the aggregators. However, the protocol is vulnerable if any intermediate node and any one (or more) of its children are compromised due to its failure to verify authenticity of the nodes because of delayed disclosure of the symmetric keys.

Cam et al. proposed an *energy-efficient secure patter-based data aggregation* (ESPDA) protocol for WSNs [11]. ESPDA is applicable for hierarchical WSNs. In ESPDA, a designated cluster-head node sends requests to sensor nodes for sending pattern codes of the data being sensed. If multiple sensor nodes send the same pattern code to the cluster-head, only one of the sensor nodes is allowed to send its data. ESPDA is a secure protocol since the aggregator nodes are not required to decrypt the data for aggregation. However, it has a high computational overhead.

To defend against attacks by malicious aggregator nodes in WSNs which may falsely manipulate data during the aggregation process, a mechanism using cryptographic technique has been proposed by Wu et al. [12]. An aggregation tree is constructed that monitors the activities of the aggregator nodes. The child nodes of an aggregator monitor the incoming data to the aggregator using the *neighborhood watch* mechanism [13], and invoke a voting protocol if any suspicious activity by the aggregator node is observed.

Ozdemir has proposed a secure and reliable data aggregation scheme that makes use of *web of trust* among the nodes in a WSN [14]. The concepts of trust and reputation have been extensively used for designing security protocols for multi-hop wireless networks like *mobile ad hoc networks* (MANETs), *wireless mesh networks* (WMNs) and WSNs [15, 16, 17]. In the scheme proposed by Ozdemir, sensor nodes exchange trust values in their neighborhood to form a web of trust that facilitates in determining secure and reliable paths to the aggregator nodes. The sensor nodes which belong to the web of trust are more reliable and their observations are given higher weights in the data aggregation process.

An aggregation protocol based on a distributed estimation algorithm for WSNs has been proposed in [18]. The protocol is secure and resistant to insider attack by any malicious or compromised node in the network.

Over the past few years, several schemes have been proposed in the literature for privacy-preserving data aggregation in WSNs. One of the most elegant approaches in this regard is homomorphic encryption [19]. Westhoff et al. have proposed *additive homomorphic encryption* scheme (it allows addition operation to be carried out on encrypted input data without requiring decryption of the cipher-text inputs) to ensure end-to-end data privacy of the sensor nodes and the base station [7]. Armknecht et al. have presented a symmetric encryption scheme for WSN data aggregation [8]. The scheme is a *bi-homomorphic encryption* scheme since it is homomorphic both for data and the keys for additive functions. Castellucia et al. proposed a scheme that utilizes an inexpensive encryption technique in aggregation methods to ensure data privacy in WSNs. A very simple, yet elegant approach for privacy-preserving multi-party computation is proposed by Chaum [20]. Castellucia et al. have described how inexpensive encryption operations may be exploited in aggregation schemes for WSNs to ensure data privacy [5].

## III. THE CPDA SCHEME FOR DATA AGGREGATION

### A. The Network Model

The basic idea of CPDA is to introduce noise to the raw data sensed from a WSN, such that an aggregator can obtain accurate aggregated information but not individual data points [3]. This is similar to the data perturbation approach extensively used in privacy-preserving data mining. However, unlike in privacy-preserving data mining where noises are independently generated (at random) and therefore leads to imprecise aggregated results, the noises in CPDA are carefully designed to leverage the cooperation between different sensor nodes, such that the precise aggregated values can be obtained by the aggregator. In particular, CPDA classifies sensor nodes into two categories: cluster leaders and cluster members. There is a one-to-many mapping between the cluster leaders and cluster members. The cluster leaders are responsible for directly aggregating data from cluster members, with the communication secured by a different shared key between any pair of communicating nodes.

The WSN is modeled as a connected graph *G(V, E)*, where *V* represents the set of senor nodes and *E* represents the set of wireless links connecting the sensor nodes. The number of sensor nodes is taken as |V| = N.

A data aggregation function is taken that aggregates the individual sensor readings. CPDA scheme has focused on additive aggregation function, $f(t) = \sum_{i=1}^{N} d_i(t)$, where $d_i(t)$ is the individual sensor reading at time instant to for node i. For computation of the aggregate functions, the following requirements are to be satisfied: (i) privacy of the individual sensor data is to be protected, i.e., each node's should be known none other expect the node itself, (ii) the number of messages transmitted within the WSN for the purpose of data aggregation should be kept at a minimum, and (iii) the aggregation result should be as accurate as possible.

### B. Key Distribution and Management

CPDA uses a random key distribution mechanism proposed in [21] for encrypting messages to prevent message eavesdropping attacks. The key distribution scheme has three phases: (i) key pre-distribution, (ii) shared-key discovery, and (iii) path-key establishment. These phases are described briefly as follows.

A large key-pool of *K* keys and their identities are first generated in the key pre-distribution phase. For each sensor nodes, *k* keys out of the total *K* keys are chosen. These *k* keys form a key ring for the sensor node.

During the key-discovery phase, each sensor node identifies which of its neighbors share a common key with itself by invoking and exchanging discovery messages. If a pair of neighbor nodes share a common key, then it is possible to establish a secure link between them.

In the path-key establishment phase, an end-to-end path-key is assigned to the pairs of neighboring nodes who do not share a common key but can be connected by two or more multi-hop secure links at the end of the shared-key discovery phase. At the end of the key distribution phase, the probability that any pair of nodes possess at least one common key is given by (1):

$$p_{connect} = 1 - \frac{((K-k)!)^2}{(K-2k)!K!} \qquad (1)$$

If the probability that any other node can overhear the encrypted message by a given key is denoted as $p_{overhear}$, then $p_{overhear}$ is given by (2):

$$p_{overhear} = \frac{k}{K} \qquad (2)$$

It has been shown in [3] that the above key distribution algorithm is efficient for communication in a large-scale sensor networks, when there is a limited number of keys available for encryption of the messages to prevent eavesdropping attacks.

### C. Cluster-Based Private Data Aggregation (CPDA)

The CPDA scheme works in three phases: (i) cluster formation, (ii) computation of aggregate results in clusters, and (ii) cluster data aggregation. These phases are described below.

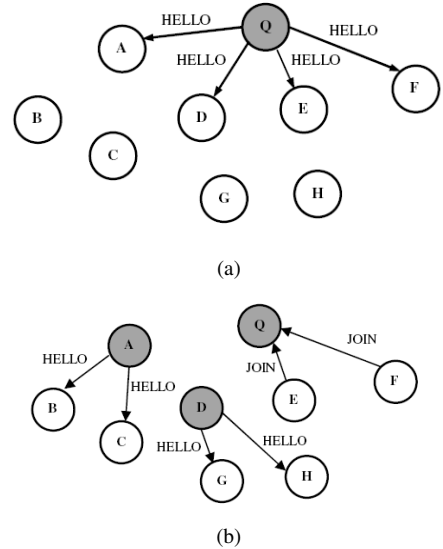

Figure 1. (a) Query server Q sends HELLO message to its neighbors. (b) An and D randomly elect themselves as the cluster leaders and send HELLO messages to their neighbors. E and F join the cluster formed by Q. B and C join the cluster formed with A as the cluster leader.

*Cluster formation*: Fig. 1 depicts the cluster formation process. A query server *Q* triggers a query by a *HELLO* message. When the *HELLO* message reaches a sensor node, it elects itself as a cluster leader with a pre-defined probability $p_c$. If a node becomes a cluster leader, it forwards the *HELLO* message to its neighbors; otherwise, it waits for a threshold period of time to check whether any *HELLO* message arrives at it from any of its neighbors. If any *HELLO* message arrives at the node, it decides to join the cluster formed by its neighbor be broadcasting a *JOIN* message. This process is repeated and multiple clusters are formed so that the entire WSN becomes a collection of a set of clusters.

*Computation within clusters*: In this phase, aggregation is done in each cluster. The computation is illustrated with the example of a simple case where a cluster contains three members: A, B, and C, where A is the assumed as the cluster leader and the aggregator, B and C are the cluster members. Let a, b, c represent the private data held by the nodes A, B, and C respectively. The goal of the aggregation scheme is to compute the sum of a, b and c without revealing the private values of the nodes.

As shown in Fig. 2, for the privacy-preserving additive aggregation function, the nodes A, B, and C are assumed to share three public non-zero distinct numbers, which are denoted as $x$, $y$, and $z$ respectively. In addition, node A generates two random numbers $r_1^A$ and $r_2^A$, which are known only to node A. Similarly, nodes B and C generate $r_1^B$, $r_2^B$ and $r_1^C$, $r_2^C$ respectively.

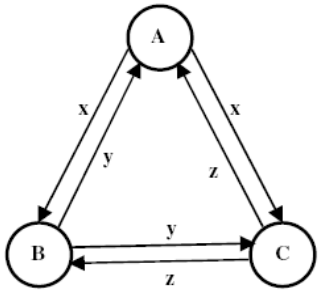

Figure 2. Nodes A, B and C broadcast their public seeds x, y and z respectively

Node A computes $v_A^A$, $v_B^A$, and $v_C^A$ as represented in (3):

$$\left.\begin{array}{l} v_A^A = a + r_1^A x + r_2^A x^2 \\ v_B^A = a + r_1^A y + r_2^A y^2 \\ v_C^A = a + r_1^A z + r_2^A z^2 \end{array}\right\} \quad (3)$$

Similarly, node B computes $v_A^B$, $v_B^B$, $v_C^B$ using (4):

$$\left.\begin{array}{l} v_A^B = b + r_1^B x + r_2^B x^2 \\ v_B^B = b + r_1^B y + r_2^B y^2 \\ v_C^B = b + r_1^B z + r_2^B z^2 \end{array}\right\} \quad (4)$$

Likewise, node C computes $v_A^C$, $v_B^C$, and $v_C^C$ using (5):

$$\left.\begin{array}{l} v_A^C = c + r_1^C x + r_2^C x^2 \\ v_B^C = c + r_1^C y + r_2^C y^2 \\ v_C^C = c + r_1^C z + r_2^C z^2 \end{array}\right\} \quad (5)$$

Node A encrypts $v_B^A$ and sends to B using the shared key between A and B. It also encrypts $v_C^A$ and sends to C using the sharing key between A and C. In the same manner, node B sends encrypted $v_A^B$ to A and $v_C^B$ to node C; node C sends encrypted $v_A^C$ and $v_B^C$ to nodes A and B respectively. The exchange of these encrypted messages is depicted in Fig. 3. On receiving $v_A^B$ and $v_A^C$, node A computes the sum of $v_A^A$ (already computed by A), $v_A^B$ and $v_A^C$. Now, node A computes $F_A$ using (6) as follows:

$$F_A = v_A^A + v_A^B + v_A^C = (a+b+c) + r_1 x + r_2 x^2 \quad (6)$$

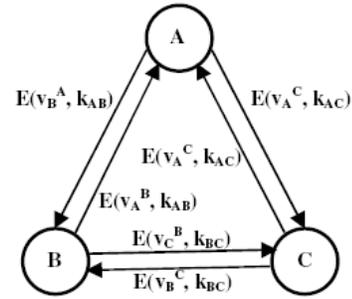

Figure 3. Exchange of encrypted messages among nodes A, B and C using their shared keys

In (6), $r_1 = r_1^A + r_1^B + r_1^C$ and $r_2 = r_2^A + r_2^B + r_2^C$. Similarly, nodes B and C compute $F_B$ and $F_C$ respectively, where $F_B$ and $F_C$ are given by (7) and (8) respectively:

$$F_B = v_B^A + v_B^B + v_B^C = (a+b+c) + r_1 y + r_2 y^2 \quad (7)$$

$$F_C = v_C^A + v_C^B + v_C^C = (a+b+c) + r_1 z + r_2 z^2 \quad (8)$$

The nodes B and C broadcast $F_B$ and $F_C$ to the cluster leader node A, so that the node A has the knowledge of the values of $F_A$, $F_B$ and $F_C$. From these values the cluster leader A can compute the aggregated value $(a + b + c)$ as explained below.

The equations (6), (7), and (8) can be rewritten as in (9):

$$U = G^{-1} F \quad (9)$$

where, $G = \begin{bmatrix} 1 & x & x^2 \\ 1 & y & y^2 \\ 1 & z & z^2 \end{bmatrix}$, $U = \begin{bmatrix} a+b+c \\ r_1 \\ r_2 \end{bmatrix}$, and

$F = \begin{bmatrix} F_A & F_B & F_c \end{bmatrix}^T$.

Since $x$, $y$, $z$, $F_A$, $F_B$, and $F_C$ are known to the cluster leader A, it can compute the value of $(a + b + c)$, without having any knowledge of b and c.

In order to avoid eavesdropping attack by neighbors, it is necessary to encrypt the values of $v_B^A$, $v_C^A$, $v_A^B$, $v_C^B$, $v_A^C$, and $v_B^C$. If node B overhears the value of $v_C^A$, then node B gets access to the values of $v_C^A$, $v_B^A$ and $F_A$. Then node B can deduce $v_A^A = F_A - v_A^B - v_A^C$. Having the knowledge of $v_A^A$, the node B can further obtain the value of a if x, $v_A^A$, $v_A^B$ and $v_A^C$ are known. However, if node A encrypts $v_C^A$ and sends it to node C, then node B cannot get $v_C^A$. With the knowledge of $v_B^A$, $F_A$ and $x$ from node A, node B cannot deduce the value of $a$. However, if nodes B and C collude and reveal A's information ($v_B^A$ and $v_C^A$), to each other, then A's privacy will be compromised and its private value $a$ will be revealed. In order to reduce the probability of such attacks, the cluster size should be as large as possible, since in a large cluster (of size $m$), at least ($m$- 1) nodes should collude in order to successfully launch the attack.

### D. Cluster Data Aggregation

The CPDA scheme has been implemented on top of a protocol known as *Tiny Aggregation* (TAG) protocol [22]. Using the TAG protocol, each cluster leader node routes the sum of values in its cluster to the query server through a TAG routing tree whose root is situated at the server.

## IV. AN ATTACK ON THE CPDA PROTOCOL

In this section, we present an efficient attack on the CPDA aggregation scheme. The objective of the attack is to show the vulnerability of the CPDA scheme which can be suitably exploited by a malicious participating sensor node. The intention of the malicious node is to participate in the scheme in such a way that it can get access to the private values (i.e., $a$, $b$ and $c$) of the participating sensor nodes. For describing the attack scenario, we use the same example cluster consisting of three sensor nodes $A$, $B$ and $C$. The node $A$ is the cluster leader whereas $B$ and $C$ are the cluster members. We distinguish two types of attacks: (i) attack by a malicious cluster leader (e.g., node $A$) and (ii) attack by a malicious cluster member (e.g., either node $B$ or node $C$). These two cases are described in detail in the following sub-sections.

### A. Attack by a Malicious Cluster Leader Node

Let us assume that the cluster leader $A$ is malicious. Node $A$ chooses a very large value of $x$ so that $x >> y, z$. Since the values of $y$ and $z$ are public values of $B$ and $C$ which are broadcast by their sources (i.e., $B$ and $C$), it is easy for $A$ to choose a suitable value for $x$.

Nodes $A$, $B$ and $C$ compute the values of $v_A^A$, $v_B^A$, $v_C^A$, $v_A^B$, $v_B^B$, $v_C^B$, $v_A^C$, $v_B^C$, and $v_C^C$ using (3), (4) and (5) as described in Section III C. As per the CPDA scheme, node $A$ receives $v_A^B = b + r_1^B x + r_2^B x^2$ from node $B$. Since $x$ is very large compared to $b$ and $r_1^B$, node $A$ can derive the value of $r_2^B$ using (10) as follows:

$$v_A^B / x^2 = b/x^2 + r_1^B / x + r_2^B = r_2^B \qquad (10)$$

Using the value of $r_2^B$ as derived in (10), node $A$ can now compute the value of $r_1^B$ using (11) as follows:

$$(v_A^B - r_2^B x^2)/x = b/x + r_1^B = r_1^B \qquad (11)$$

In the same manner, node $A$ derives the values of $r_1^C$ and $r_2^C$ from $v_A^C$ that it receives from node $C$. Since $r_1 = r_1^A + r_1^B + r_1^C$, and $r_2 = r_2^A + r_2^B + r_2^C$, in (6), (7) and (8), node $A$ can compute the values of $r_1$ and $r_2$ ($r_1^B$, $r_2^B$, $r_1^C$, and $r_2^C$ are derived and $r_1^A$ and $r_2^A$ are generated by node A).

At this stage, node $A$ uses $F_B$ and $F_C$ values as sent by nodes B and C in (7) and (8) respectively. Node A has now two linear simultaneous equations with two unknowns: $b$ and $c$, the values of $y$ and $z$ being public. Solving (7) and (8) for $b$ and $c$, the malicious cluster leader node $A$ can get the access to the private information of nodes $B$ and $C$, thereby launching a privacy attack on the CPDA scheme.

### B. Attack by a Malicious Cluster Member Node

In this scenario, let us assume that the cluster member $B$ is malicious and it tries to access the private values of the cluster leader $A$ and the cluster member $C$. Node $B$ chooses a very large value of $y$ so that $y >> x, z$. Once the value of $F_B$ is computed in (7), node $B$ derives the value of $r_2$ and $r_1$ using (12) and (13) as follows:

$$F_B / y^2 = (a+b+c)/y^2 + r_1/y + r_2 = r_2 \qquad (12)$$

$$(F_B - r_2 y^2)/y = (a+b+c)/y + r_1 = r_1 \qquad (13)$$

As per the CPDA scheme, node $B$ receives $v_B^C = c + r_1^C y + r_2^C y^2$ from node C. Since $y$ is very large compared to $c$ and $r_1^C$, and $r_2^C$ node $B$ can derive the value of $r_2^C$ and $r_1^C$ using (14) and (15) as follows:

$$v_B^C / y^2 = c/y^2 + r_1^C / y + r_2^C = r_2^C \qquad (14)$$

$$(v_B^C - r_2^C y^2)/y = c/y + r_1^C = r_1^C \qquad (15)$$

Using (12), (13), (14) and (15), node $B$ can compute $r_1^A = r_1 - r_1^B - r_1^C$, and $r_2^A = r_2 - r_2^B - r_2^C$. Now, node $B$ can compute the value of $a$ using $v_B^A = a + r_1^A y + r_2^A y^2$ (received from A), in which values of all the variables are known except $a$. In a similar fashion, node $B$ derives the value of $c$ using $v_B^C = c + r_1^C y + r_2^C y^2$ (received from C).

Since the private values of the nodes $A$ and $C$ are known to node $B$, the privacy attack launched by participating cluster member node $B$ is successful on the CPDA aggregation scheme.

## V. PROPOSED MODIFICATION IN CPDA PROTOCOL

In this section, we present two modifications of CPDA scheme: one towards making the protocol more efficient and the other for making it more secure and robust.

### A. Modified CPDA for Enhanced Energy Efficiency

In this section, a modification is proposed for the CPDA protocol for achieving enhanced efficiency of its operation. The modification is based on suitable choice for the value of *x* (the public seed) done by the aggregator node *A*.

Let us assume that the node *A* chooses a large value of *x* such that the following conditions are satisfied.

$$r_2 x^2 \gg r_1 x \qquad (16)$$
$$r_1 x \gg (a+b+c) \qquad (17)$$

In (16) and (17), $r_1 = r_1^A + r_1^B + r_1^C$ and $r_2 = r_2^A + r_2^B + r_2^C$. Now, node *A* has computed the value of $F_A$ as in (6). In order to efficiently compute the value of $(a + b + c)$, node *A* divides the value of $F_A$ by $x^2$ as shown in (18)

$$F_A / x^2 = (a+b+c)/x^2 + r_1 x / x^2 + r_2 = r_2 \qquad (18)$$

Using (18), node *A* derives the value of $r_2$. Once the value of $r_2$ is deduced, node *A* attempts to compute the value of $r_1$ as follows.

$$F_A - r_2 x^2 = (a+b+c) + r_1 x \qquad (19)$$
$$r_1 = (F_A - r_2 x^2)/x - (a+b+c)/x = (F_A - r_2 x^2)/x \qquad (20)$$

Since, the values of $F_A$, $r_2$ and *x* are all known to node *A*, it can compute the value of $r_1$ using (13). Once the values of $r_1$ and $r_2$ are computed by node *A*, it can compute the value of $(a + b + c)$ using (6). Since the computation of the sum $(a + b + c)$ by node *A* involves two division operations only (as done in (12) and (13)), the modified CPDA scheme will be extremely light-weight and hence much more energy- and time- efficient as compared with the original CPDA scheme. The original CPDA scheme involved extra computation of the values of $F_B$ and $F_C$ an expensive matrix inversion operation as described in Section III C.

### B. Modified CPDA for Increased Security and Robustness

In this section, we discuss the modifications required on the existing CPDA scheme so that a malicious participant node cannot launch the attack described in Section IV.

It may be noted that, the essential vulnerability of the CPDA scheme lies in the unrestricted freedom delegated on the participating nodes for generating their public seed values. For example, nodes *A*, *B* and *C* have no restrictions on values of *x*, *y* and *z* respectively while these values are generated by the nodes. A malicious attacker exploits this freedom to generate an arbitrarily large public seed value and launches the attack as discussed in Section IV.

In order to prevent such an attack, the CPDA protocol needs to be modified. In this modified version, the nodes in a cluster make a check on the generated public seed values so that it is not possible for a malicious participant to generate any arbitrarily large seed value. For a cluster with three nodes, such a constraint may be imposed by the requirement that the sum of any two public seeds must be greater than the third seed. In other words: $x + y > z$, $z + x > y$, and $y + z > x$. If these constraints are satisfied by the generated values of *x*, *y* and *z*, it will be impossible for any node to launch the attack and get access to the private values of the other participating nodes.

However, even if the above restrictions on the values of *x*, *y* and *z* are imposed, the nodes should be careful in choosing the values for their secret random number pairs. If two nodes happen to choose very large values for their random numbers compared to those chosen by the third node, then it will be possible for the third node to get access to the private values of the other two nodes. For example, let us assume that nodes *A* and *C* have chosen the values of $r_1^A$, $r_2^A$ and $r_1^C$, $r_2^C$ such that they are all much larger than $r_1^B$ and $r_2^B$ - the private random number pair chosen by node *B*. It will be possible for node *B* to derive the values of *a* and *c*: the private values of nodes *A* and *C* respectively. This is explained in the following.

Node *B* receives $v_B^A = a + r_1^A y + r_2^A y^2$ from node *A* and computes the values of $r_1^A$ and $r_2^A$ using (21) and (22) as follows:

$$v_A^A / y^2 = a/y^2 + r_1^A / y + r_2^A = r_2^A \qquad (21)$$
$$(v_B^A - r_2^A y^2)/y = a/y + r_1^A = r_1^A \qquad (22)$$

In a similar fashion, node *B* derives the values of $r_1^C$ and $r_2^C$ from $v_B^C$ received from node *C*. Now, the node *B* computes $r_1 = r_1^A + r_1^B + r_1^C$ and $r_2 = r_2^A + r_2^B + r_2^C$ since it has access to the values of all these variables. The values of $F_B$ and $F_C$ are broadcast by nodes *B* and *C* in unencrypted from. Hence, the node *B* has access to both these values. Using (7) and (8), node *B* computes the values of *a* and *c*, since these are the two unknowns in the two linear simultaneously equations.

In order to defend against the above vulnerability, the CPDA protocol needs further modification. In this modified version, after the values $v_A^A$, $v_A^B$, and $v_A^C$ are generated and shared by *A*, *B* and *C* respectively, the nodes check whether the following constraints are satisfied: $v_A^A + v_A^B > v_A^C$, $v_A^B + v_A^C > v_A^A$, and $v_A^C + v_A^A > v_A^B$. The nodes proceed for further execution of the algorithm only if the above three inequalities are satisfied. If all three inequalities are not satisfied, there will be a possibility that the random numbers generated by one node is much larger than those generated by other nodes – a scenario which indicates a possible attack by a malicious node.

## VI. CONCLUSION AND FUTURE WORK

In-network data aggregation in WSNs is a technique that combines partial results at the intermediate nodes en route to the base station (i.e. the node issuing the query), thereby reducing the communication overhead and optimizing the bandwidth utilization in the wireless links. However, this techniques raises privacy issues of the sensor nodes which needs to share their data with the aggregator node. In applications such as health care and military surveillance where the sensitivity of the private data of the sensors is very high, the aggregation has to be carried out in a privacy-preserving way, so that the sensitive data are not revealed to the aggregator. A very popular scheme for this purpose exists in the literature which is known as CPDA. Although, CPDA is in literature for quite some time now, no vulnerability of the protocol has been identified so far. In this paper, we have first shown a security vulnerability in the CPDA protocol, in which we have demonstrated how a malicious sensor node in a WSN can exploit the protocol in such a way that it gets access to the private sensitive values of its neighboring nodes while data aggregation process takes place in an aggregator. We have also proposed a suitable modification of the CPDA protocol to make it robust against this vulnerability and also to make it computationally more efficient. Future plan of work includes experimental analysis to evaluate the performance of the proposed modified CPDA protocol and compare its computational and communication overhead with those of the existing privacy homomorphism-based encryption for secure data aggregation in WSNs.